\hfuzz 2pt
\font\titlefont=cmbx10 scaled\magstep1
\magnification=\magstep1

\null
\vskip .5cm
\centerline{\titlefont NEUTRAL KAONS IN RANDOM MEDIA}
\vskip 2.5cm
\centerline{\bf F. Benatti}
\smallskip
\centerline{Dipartimento di Fisica Teorica, Universit\`a di Trieste}
\centerline{Strada Costiera 11, 34014 Trieste, Italy}
\centerline{and}
\centerline{Istituto Nazionale di Fisica Nucleare, Sezione di 
Trieste}
\vskip 1cm
\centerline{\bf R. Floreanini}
\smallskip
\centerline{Istituto Nazionale di Fisica Nucleare, Sezione di 
Trieste}
\centerline{Dipartimento di Fisica Teorica, Universit\`a di Trieste}
\centerline{Strada Costiera 11, 34014 Trieste, Italy}
\vskip 1cm
\centerline{\bf R. Romano}
\smallskip
\centerline{Dipartimento di Fisica Teorica, Universit\`a di Trieste}
\centerline{Strada Costiera 11, 34014 Trieste, Italy}
\centerline{and}
\centerline{Istituto Nazionale di Fisica Nucleare, Sezione di 
Trieste}
\vskip 2.5cm
\centerline{\bf Abstract}
\smallskip
\midinsert
\narrower\narrower\noindent
The generalized dynamics describing the propagation of neutral kaons
in randomly fluctuating media is derived and analyzed. It takes into
account possible matter-induced effects leading to loss of phase
coherence and dissipation. The study of selected neutral kaon observables
indicates that these non-standard effects are amenable to a direct
experimental analysis.
\endinsert
\bigskip
\vfill\eject

{\bf 1. INTRODUCTION}
\medskip

When a quantum system is immersed in an external environment,
its time evolution is no longer unitary, since in general it
exchanges energy and entropy with the larger system.
In principle, the subdynamics can be readily obtained from the total,
unitary system+environment evolution by eliminating
({\it i.e.} integrating over) the degrees of freedom pertaining to the
environment. In practice however, this procedure leads to tractable
dynamical equations only when the interaction between subsystem and 
environment can be considered weak: in these cases memory 
effects disappear, and the time evolution of the subsystem
is described by linear maps, encoding non-hamiltonian physical effects,
like irreversibility and dissipation.[1-4]

The system plus environment paradigm for the
treatment of the so-called open systems is nevertheless very general
and has been successfully adopted to model very different physical
situations, in laser, atomic and molecular physics.[1-8]
In particular, it has been very useful in describing the effects
of random media or of stochastic external fields in particle
propagation inside interferometric devices.[9-13] Indeed,
for weakly-coupled systems, the decoherence and dissipative phenomena
induced by the media are in general very small, so that the most
suited way to study them is through appropriate interferometric set-ups.[14-16]

Motivated by these results, in the following similar techniques will
be adopted to study effects of irreversibility and dissipation
in neutral kaons propagating inside a random medium.
Because of strangeness oscillations, neutral kaons have been one of the
prime laboratory for the discovery and analysis of small physical effects,
$CP$ violation being the most striking example;[17,18] therefore, they appear
to be the natural place for studying environment-induced dissipative effects,
also in view of the unique opportunities offered by the
production of correlated kaons at $\phi$-factories.[19]

Matter effects in the physics of neutral kaons have been studied since
the very early days of kaon physics. However, most of the attention
has been devoted to the so called regeneration phenomena, where
short-lived neutral kaon states are regenerated in a beam of purely 
long-lived ones via the passage through a thin slab of material:
this is due to the coherent interactions of the kaons with the
nuclei of the material, that produce different scattering amplitudes
for the kaon and antikaon components of the impinging beam.
One can show that this effect is dominated by forward scattering
processes, and it can be described by an effective hamiltonian
involving the index of refraction of the material.[20-23]

Here, we shall generalize this physical situation by considering
the propagation of neutral kaons in a randomly fluctuating medium.
The kaon system can then be viewed as an open system, where the
environment (the medium) is described by a classical, random 
external field.

Quite in general, any environment can be modelled in this way,
provided the characteristic decay time of the associated correlations
is sufficiently small with respect to the typical evolution time
$\tau_S$ of the subsystem. In the case of the kaons, $\tau_S$
can be roughly identified with the lifetime of the short-lived kaon,
so that the correlations in the material through which the kaons
propagate must decay very rapidly, or equivalently, the medium
must fluctuate on times much shorter than $\tau_S$.
Although this condition looks quite restrictive, it can be met
quite easily by a careful choice of material. Indeed, many
short-time physical phenomena, like molecular vibrational motion
and relaxation, or collisions in liquids, take place at times 
that are at least a couple of orders of magnitude smaller
than $\tau_S$.%
\footnote{$^{1)}$}{The direct study of these very short-lived 
phenomena have been made possible by the recent introduction of lasers 
that are able to produce pulses reaching the femtosecond scale;
for a review, see [24].}
Therefore, a kaon moving {\it e.g.} in a gas at sufficiently high
temperature or in a liquid would see a random fluctuating distribution
of centers of scattering, and therefore be subjected to stochastic
incoherent interactions with the material; as a consequence,
its dynamics can be effectively modelled as being that of a particle
propagating in a random medium.

This physical situation is clearly different from the one encountered in
the regeneration phenomena, where a single coherent scattering in the
thin slab is enough to account for the effect. In the present case instead,
correlations in the material play a fundamental role: as we shall see,
they are responsible for the generation of irreversibility and 
loss of quantum coherence.%
\footnote{$^{2)}$}{Using a more phenomenological approach,
non-standard effects in the propagation
of neutral kaons in a stationary, low-density 
material have been recently discussed in [25].}

In the next section, we shall discuss in detail the derivation of 
the master equation that describes within the weak-coupling hypothesis
the propagation of neutral kaons in random media. Since it incorporates
the presence of dissipative phenomena, it can not be fully written
in the familiar hamiltonian form; 
rather, it assumes the structure appropriate for generating a quantum
dynamical semigroup.[1-4] Specific examples are presented in Sect.3, while
some general properties of its solutions are discussed in Sect.4.
Since the matter induced effects are small, a suitable perturbative
approximation can be used, allowing the explicit evaluation of
relevant kaon observables: Sect.5 and 6 will be devoted to
the analysis of the behaviour of these observables.
We shall first study the decays of single kaons, relevant for
fixed-target experiments, and then discuss the case of correlated
kaons at $\phi$-factories. In both cases, the matter-induced phenomena
modify in a very specific way the various kaon observables, that are 
therefore amenable to a direct experimental analysis.
In particular as discussed in the final Sect.7,
at $\phi$-factories one can adjust the experimental
conditions so as to allow the propagation in the medium of only
one of the two correlated kaons; in this configuration, one can
easily perform tests on the physical consistency of the adopted treatment,
shedding further light on the general description of 
open quantum systems.

Although preliminary, we hope that the results presented in our
investigation will stimulate further studies on the dynamics
of neutral kaons in random media, encouraging in 
particular a detailed experimental analysis.

\vfill\eject

{\bf 2. MASTER EQUATION}
\medskip

We shall work within the familiar effective description of the neutral 
kaon system which requires the introduction of a two-dimensional Hilbert
space;[17-19] the set $\{|K^0\rangle, |\overline{K^0}\rangle\}$ constitutes
a convenient basis in this space. With respect to this basis,
single kaon states can then be represented by density matrices,
$\widehat{R}$, {\it i.e.} by hermitian $2\times2$ matrices with non-negative
eigenvalues. Their time evolution is described in terms of an equation
of a standard Liouville-von Neumann form:
$$
{\partial\widehat{R}(t)\over\partial t}=-iH^{(0)}\, \widehat{R}(t)
+i\widehat{R}(t)\, H^{(0)}{}^\dagger 
+L_t\big[\widehat{R}(t)\big] .\eqno(2.1)
$$
The first two pieces on the r.h.s. give the usual hamiltonian contribution,
while the additional linear map $L_t$, explicitly given in (2.10) below,
takes into account the presence of the stochastic medium.

The effective hamiltonian $H^{(0)}$, the so-called 
Weisskopf-Wigner hamiltonian,
describes the propagation of the kaons in vacuum;%
\footnote{$^{3)}$}{Throughout the paper, we shall append a
superscript $^{(0)}$ to all quantities referring to neutral kaons
in vacuum, {\it i.e.} in absence of the medium.}
it contains a non-hermitian term, characterized by the natural width 
of the physical states. The entries of $H^{(0)}$ can be expressed
in terms of its complex eigenvalues,
$\lambda_S^{(0)}=m_S^{(0)}-i\gamma_S^{(0)}/2$, 
$\lambda_L^{(0)}=m_L^{(0)}-i\gamma_L^{(0)}/2$,
and the complex parameters $p_S^{(0)}$, $q_S^{(0)}$, $p_L^{(0)}$, $q_L^{(0)}$, 
appearing in the corresponding (right) eigenstates, 
$$
\eqalign{
&|K_S\rangle=p_S^{(0)}\, |K^0\rangle + q_S^{(0)}\, |\overline{K^0}\rangle\ ,\qquad\quad
\big|p_S^{(0)}\big|^2 + \big|q_S^{(0)}\big|^2=1\ ,\cr
&|K_L\rangle=p_L^{(0)}\, |K^0\rangle - q_L^{(0)}\, |\overline{K^0}\rangle\ ,\qquad\quad
\big|p_L^{(0)}\big|^2 + \big|q_L^{(0)}\big|^2=1\ .}
\eqno(2.2)
$$
For later convenience, it is useful to introduce the following positive
combinations, involving the eigenstates masses and widths,
$$
\Delta m^{(0)}=m_L^{(0)} - m_S^{(0)}\ ,\qquad
\Delta\Gamma^{(0)}=\gamma_S^{(0)}-\gamma_L^{(0)}\ ,
\qquad \Gamma^{(0)}={\gamma_S^{(0)}+\gamma_L^{(0)}\over2}\ ,
\eqno(2.3)
$$
as well as the complex quantities:
$$
\Gamma_\pm^{(0)}=\Gamma^{(0)}\pm i \Delta m^{(0)}\ ,\qquad
\Delta\Gamma_\pm^{(0)}=\Delta\Gamma^{(0)}\pm 2i \Delta m ^{(0)}\ .
\eqno(2.4)
$$
The effective Hamiltonian $H^{(0)}$ can be diagonalized using the
similarity transformation induced by (2.2):
$$
H^{(0)}=V^{(0)}\, H_D^{(0)}\, V^{(0)}{}^{-1}\ ,
\eqno(2.5)
$$
with
$$
V^{(0)}=\left[\matrix{p_S^{(0)}&\phantom{-}p_L^{(0)}\cr
                q_S^{(0)}&-q_L^{(0)}}\right]\ ,\ \qquad
H_D^{(0)}=\left[\matrix{\lambda_S^{(0)}&0\cr
                0&\lambda_L^{(0)}}\right]\ .
\eqno(2.6)
$$
Then, one can write:
$$
H^{(0)}={\lambda_S^{(0)}+\lambda_L^{(0)}\over 2}
+{\lambda_S^{(0)}-\lambda_L^{(0)}\over 2}\
\left[\matrix{\theta^{(0)}&{2\sigma^{(0)}\over r_S^{(0)}+r_L^{(0)}}\cr
                     {2\over r_S^{(0)}+r_L^{(0)}}&-\theta^{(0)}}\right]\ ;
\eqno(2.7)
$$
the two complex parameters
$$
\theta^{(0)}={r_S^{(0)}-r_L^{(0)}\over r_S^{(0)}+r_L^{(0)}}\ ,
\qquad\quad
\sigma^{(0)}=r_S^{(0)}\, r_L^{(0)}\ ,
\eqno(2.8)
$$
involving the ratios,
$$
r_S^{(0)}={p_S^{(0)}\over q_S^{(0)}}\ ,\ \qquad 
r_L^{(0)}={p_L^{(0)}\over q_L^{(0)}}\ ,
\eqno(2.9)
$$
signal $CPT$ (for $\theta^{(0)}\neq0$) and $T$
(for $\xi^{(0)}\equiv(|\sigma^{(0)}|-1)/(|\sigma^{(0)}|+1)\neq0$)
violating effects in mixing, respectively.

All this is valid in vacuum; when the neutral kaons propagate in matter,
the interaction with the nuclei of the material gives additional contributions
to the evolution equation, the term $L_t$ in (2.1). As explained in the
Introduction, we shall consider the case of rapidly fluctuating media,
that can be represented by classical stochastic fields.
The action of the media on the kaons can then be
expressed in generalized hamiltonian form:
$$
L_t\big[\widehat{R}(t)\big]=-i F(t)\, \widehat{R}(t) 
+ i \widehat{R}(t)\, \big[F(t)\big]^\dagger\ ,
\eqno(2.10)
$$
where
$$
F(t)=\sum_{\mu=0}^3 F_\mu(t)\, \sigma_\mu\ ,
\eqno(2.11)
$$
while $\sigma_0$ is the $2\times2$ unit matrix
and $\sigma_1$, $\sigma_2$, $\sigma_3$ the Pauli matrices.
The quantities $F_\mu(t)$, $\mu=0,1,2,3$, generate a complex, 
Gaussian stochastic field;
they are assumed to have in general non-zero mean, 
but translationally invariant correlations
(a star means complex conjugation):
$$
\eqalignno{
&\widehat{G}_{\mu\nu}(t-s)\equiv\langle F_\mu(t)\, F_\nu(s)\rangle
-\langle F_\mu(t)\rangle\, \langle F_\nu(s)\rangle\ , &(2.12a)\cr
&\widehat{W}_{\mu\nu}(t-s)\equiv\langle F_\mu(t)\, F_\nu^*(s)\rangle
-\langle F_\mu(t)\rangle\, \langle F_\nu^*(s)\rangle\ . &(2.12b)\cr
}
$$

Since the generalized hamiltonian $F(t)$ in (2.11) involves stochastic variables,
the density matrix $\widehat{R}(t)$, solution of the total equation of
motion (2.1), is also stochastic. Instead, we are interested in the behaviour
of the reduced density matrix $\hat{\rho}(t)\equiv\langle\widehat{R}(t)\rangle$
which is obtained by averaging over the noise; it is $\hat{\rho}(t)$ that
describes the effective evolution of the kaon states in the medium and allows
computing the behaviour of relevant observables.
We shall now explicitly describe the derivation of an effective master
equation for $\hat{\rho}(t)$, making the additional assumption that kaons
and noise be decouple at $t=\,0$, so that the initial state is
$\hat{\rho}(0)\equiv\langle\widehat{R}(0)\rangle=\widehat{R}(0)$.%
\footnote{$^{4)}$}{This condition is always satisfied in a typical
experimental situation, where the kaons enter the medium
after being produced.}

Since the the hamiltonian $H^{(0)}$ is statistically independent
from the stochastic variables, one can choose to average over the
noise in the interaction representation, where we set:
$$
\widetilde{R}(t)=e^{it\, H^{(0)}}\ \widehat{R}(t)\ e^{-it\, H^{(0)}{}^\dagger}\ ,
\eqno(2.13)
$$
so that:
$$
{\partial \widetilde{R}(t)\over\partial t}
=\widetilde{L}_t\big[\widetilde{R}(t)\big]\equiv
-i\sum_{\mu=0}^3 F_\mu(t)\, \sigma_\mu(t)\ \widetilde{R}(t)
+i\, \widetilde{R}(t)\ \sum_{\mu=0}^3 F_\mu^*(t)\, \big[\sigma_\mu(t)\big]^\dagger\ ,
\eqno(2.14)
$$
with $\sigma_\mu(t)=e^{it\, H^{(0)}}\, \sigma_\mu\ e^{-it\, H^{(0)}}$.
The time evolution of the reduced density matrix in the interaction representation,
$\tilde{\rho}(t)\equiv\langle\widetilde{R}(t)\rangle$, can then be expressed
as a series expansion involving multiple correlations
of the operator $\widetilde{L}_t$:
$$
\eqalignno{
&\tilde\rho(t)= M_t[\tilde\rho(0)]
\equiv\sum_{k=0}^\infty M_k[\tilde\rho(0)]\ , &(2.15)\cr
&M_k[\tilde\rho\,]=\int_0^t ds_1\int_0^{s_1}ds_2\cdots \int_0^{s_{k-1}}ds_k\
\langle \widetilde{L}_{s_1} \widetilde{L}_{s_2}\cdots \widetilde{L}_{s_k}\rangle
[\tilde\rho\,]\ . &(2.16)
}
$$
The sum $M_t$ in (2.15) can be formally inverted, and a suitable resummation
gives (a dot represents time derivative):[26]
$$
{\partial \tilde{\rho}(t)\over\partial t}=\dot{M}_t\, M_t^{-1}[\tilde\rho(t)]=
\Big\{\dot{M}_1+\big(\dot{M}_2-\dot{M}_1\, M_1\big)+\ldots\Big\}[\tilde\rho(t)]\ .
\eqno(2.17)
$$

As mentioned before and discussed in more detail at the end of the section,
the action of the medium on the travelling kaons is weak. Therefore,
one can focus the attention on the dominant terms of the previous expansions, 
neglecting all contributions higher than the second-order ones.
Further, since the characteristic decay time of correlations in the medium
is by assumption much smaller than the typical time scale of the system,
the memory effects implicit in (2.17) should not be physically relevant
and the use of the Markovian approximation justified. This is implemented 
in practice by extending to infinity the upper limit of the integrals
appearing in $\dot{M}_2$ and $M_1$ (compare with (2.16)).[1-4]

By returning to the Schr\"odinger representation, one finally obtains the master
equation generating the time evolution of the reduced density matrix
$\hat\rho(t)\equiv\langle\widehat{R}(t)\rangle$. 
It takes the following explicit form:
$$
{\partial \hat{\rho}(t)\over\partial t}=-iH\, \hat\rho(t)
+i\hat\rho(t)\, H^\dagger+\widehat{L}[\hat\rho(t)]\ , 
\eqno(2.18a)
$$
where
$$
\eqalignno{
&H=H^{(0)} + H^{(1)} + H^{(2)}\ , &(2.18b)\cr
&\widehat{L}[\hat\rho\,]={1\over2}\sum_{i,j=1}^3 \widehat{\cal C}_{ij}
\Big[2\sigma_i \hat\rho\, \sigma_j
- \{\sigma_j\sigma_i\, ,\, \hat\rho\}\Big]\ . &(2.18c)
}
$$
The effective hamiltonian in matter $H$ differs from the one in vacuum
$H^{(0)}$ by first order terms (coming from the piece $\dot{M}_1$
in (2.17)) involving the noise mean values:
$$
H^{(1)}=\sum_{\mu=0}^3 \langle F_\mu(t)\rangle\ \sigma_\mu\ ,
\eqno(2.19)
$$
and by second-order contributions (coming from the second-order terms in (2.17))
involving the noise correlations (2.12) through the 
time-independent combinations:
$$
\eqalignno{
&\widehat{\cal B}_{\mu\nu}=\sum_{\lambda=0}^3\int_0^\infty dt\ \widehat{G}_{\mu\lambda}(t)\ 
{\cal U}_{\lambda\nu}(-t)\ , &(2.20a)\cr
&\widehat{\cal C}_{\mu\nu}=\sum_{\lambda=0}^3\int_0^\infty dt\ 
\Big[\widehat{W}_{\mu\lambda}(t)\ {\cal U}^*_{\lambda\nu}(-t)
+\widehat{W}^*_{\nu\lambda}(t)\ {\cal U}_{\lambda\mu}(-t)\Big]\ , &(2.20b)
}
$$
where the $4\times4$ matrix ${\cal U}(t)$ is defined by the following
transformation rule:
$$
\sigma_\mu(t)=e^{it\, H^{(0)}}\ \sigma_\mu\
e^{-it\, H^{(0)}}=
{\cal U}_{\mu\nu}(t)\ \sigma_\nu\ .
\eqno(2.21)
$$
Explicitly, one finds:
$$
\eqalign{
H^{(2)}={i\over2}&\sum_{\mu=0}^3\Big(\widehat{\cal C}_{\mu\mu}
-2\widehat{\cal B}_{\mu\mu}\Big)\, \sigma_0\cr
&+\sum_{i=1}^3\bigg[i\,\widehat{\cal C}_{i0}
-i\Big(\widehat{\cal B}_{i0} + \widehat{\cal B}_{0i}\Big)
+{1\over2}\sum_{j,k=1}^3\epsilon_{ijk}\Big(2\widehat{\cal B}_{jk}
-i\,{\cal I}m\,\widehat{\cal C}_{jk}\Big)\bigg]\ \sigma_i\ .}
\eqno(2.22)
$$
On the other hand, the additional piece $\widehat{L}[\hat\rho\,]$ in $(2.18c)$
is a time-independent, trace-preserving linear map
involving the hermitian $3\times3$ submatrix $\widehat{\cal C}_{ij}$,
obtained from the coefficient matrix $(2.20b)$ by letting $\mu,\nu=1,2,3$.
It introduces irreversibility, inducing
in general dissipation and loss of quantum coherence.
Altogether, equation (2.18) generates a semigroup of linear maps,
${\mit\Gamma}_t:\ \hat\rho(0)\mapsto\hat\rho(t)\equiv{\mit\Gamma}_t[\hat\rho(0)]$,
for which composition is defined only forward in time:
${\mit\Gamma}_t\circ{\mit\Gamma}_s={\mit\Gamma}_{t+s}$, with $t,s\geq0$;
it is usually referred to as a quantum dynamical semigroup.[1-4]

As a further remark, notice that when correlations in the medium are negligible,
{\it i.e.} the combination in (2.12) are vanishingly small, 
the physical situation described by equation (2.18) corresponds to that of the
regeneration phenomenon. Indeed, in this case the presence of matter is signaled
solely by the shift $H^{(1)}$ in the effective hamiltonian.
For a kaon-medium interaction dominated by coherent forward scattering,
$H^{(1)}$ becomes diagonal and its components
turn out to be expressible in terms of the forward scattering amplitudes 
$f_K$, $f_{\overline{K}}$ in the medium:[20-23]
$$
H^{(1)}=-{2\pi\nu\over m_K}\left[\matrix{f_K & 0\cr
                                         0 & f_{\overline{K}}}\right]\ ,
\eqno(2.23)
$$
where $m_K$ is the $K^0$ mass, while $\nu$ represents the density of
scattering centers in the medium.

The amplitudes $f_K$, $f_{\overline{K}}$ have been directly measured 
in experiments for many types of materials,[27, 28] and quite consistently
one finds: $|H^{(1)}|\simeq 10^{-2}\ \Delta\Gamma^{(0)}$,
where \hbox{$\Delta\Gamma^{(0)}\simeq\gamma_S^{(0)}$} is the typical inverse timescale
describing the kaon evolution in vacuum, as generated by 
the hamiltonian $H^{(0)}$.
In view of the fact that $H^{(1)}$ coincides with the mean value of the
stochastic hamiltonian in (2.11), one concludes that also the averages
$\langle F_\mu \rangle$ should be in modulus of the same order of magnitude.
Noise correlations are however much smaller: from the definition
(2.12), one can safely deduce the following rough
estimate: $|\widehat{G}|\simeq|\widehat{W}|\simeq
|\langle F_\mu \rangle|^2\simeq 10^{-4}\ (\Delta\Gamma^{(0)})^2$.
Therefore, the weak coupling hypothesis adopted in deriving
the master equation (2.18) appears physically justified.

\vskip 1cm

{\bf 3. EXAMPLES}
\medskip

Before analyzing in detail the evolution in time of the density matrix
$\hat\rho(t)$, we shall discuss some physically interesting instances
of the master equation (2.18); they correspond to specific realizations
of the medium through which the kaons propagate.

Some general considerations apply to all cases.
As discussed at the end of the previous section, the noise contributions
to (2.18) are expected to be small, in particular those involving 
noise correlations. Therefore, whenever the correlations in (2.12)
are multiplied by other small parameters, {\it e.g.} those coming 
from the hamiltonian $H^{(0)}$ in (2.7), one can safely neglect them
in comparison with the dominant pieces. These considerations are particularly
relevant in the evaluation of $\widehat{\cal B}_{\mu\nu}$
and $\widehat{\cal C}_{\mu\nu}$ in (2.20).
These quantities are linear 
in the noise correlations: the entries of the matrix ${\cal U}(t)$
appearing in (2.20) can then be computed at lowest order.
Using the diagonalization (2.5) to write 
$e^{it\, H^{(0)}}=V^{(0)}\, e^{it\, H_D^{(0)}}\, V^{(0)}{}^{-1}$
and further setting $p_S^{(0)}=q_S^{(0)}=p_L^{(0)}=q_L^{(0)}=1/\sqrt{2}$ 
in (2.6),%
\footnote{$^{5)}$}{These conditions, that come from the assumption
of $CPT$ and $T$ (hence $CP$) conservation in mixing, imply a specific
phase choice in the definition of the basis state vectors in the Hilbert
space; in practice, this poses no problems, since physical observables,
being the result of a trace operation (see later), are manifestly
phase-invariant.}
from (2.21) one explicitly finds:
$$
{\cal U}(t)=\left[\matrix{\sigma_0 & 0\cr
                              0    & \sigma(t)}\right]\ ,\qquad
\sigma(t)=\left[\matrix{\cos\omega t & \sin\omega t\cr
                        -\sin\omega t & \cos\omega t}\right]\ ,
\eqno(3.1)
$$
where $\omega=i\Delta\Gamma_-^{(0)}/2$.

Further, the stochastic medium fluctuates on time intervals
much shorter than the typical kaon evolution timescale $1/\Delta\Gamma^{(0)}$.
Correspondingly, the noise correlations in (2.12) can be taken to have
an exponentially decreasing form, with decay parameter $\lambda$
much larger than $\Delta\Gamma^{(0)}$. This allows neglecting
all higher-order contributions in $\Delta\Gamma^{(0)}/\lambda$
while evaluating explicitly the coefficients $\widehat{\cal B}_{\mu\nu}$
and $\widehat{\cal C}_{\mu\nu}$.

\vskip .7cm

{\bf 3.1 Generalized regeneration}
\medskip

In this case, the stochastic hamiltonian in (2.11) is taken to be diagonal,
so that only the components with $\mu=0,3$ of the stochastic variables
$F_\mu(t)$ are non vanishing:
$$
F(t)=\left[\matrix{F_+(t) & 0\cr
                   0  & F_-(t)}\right]\ ,\qquad
F_\pm(t)={1\over2}\big(F_0(t)\pm F_3(t)\big)\ .
\eqno(3.2)
$$
Further, we assume the correlation functions to have the form
$(\mu,\nu=0,3)$:
$$
\eqalignno{
&\widehat{G}_{\mu\nu}(t-s)=G_{\mu\nu}\ e^{-\lambda_1 |t-s|}\ , &(3.3a)\cr
&\widehat{W}_{\mu\nu}(t-s)=W_{\mu\nu}\ e^{-\lambda_2 |t-s|}\ , &(3.3b)
}
$$
where $G_{\mu\nu}$ and $W_{\mu\nu}$ are time-independent complex matrices,
with $W=W^\dagger$. Physically speaking,
this case corresponds to a generalization of the familiar
regeneration situation. The kaons are still mainly forward scattered
by the medium; however, while travelling in it, they encounter density
fluctuations, whose correlations have the general behaviour (3.3).

With the help of (3.1) and (3.3), one can now explicitly compute
the noise contributions in (2.18). To lowest order in the small parameters,
one finds that only the $i=j=3$ entry of the matrix
$\widehat{\cal C}_{ij}$ is non vanishing,
$$
\widehat{\cal C}_{33}={2 W_{33}\over\lambda_2}\geq0\ ,
\eqno(3.4)
$$
while the hamiltonian contributions $H^{(1)}$ and $H^{(2)}$ turns out to be
diagonal:
$$
\eqalign{
H-H^{(0)}=&\bigg[\langle F_0\rangle+i{W_{00}+W_{33}\over\lambda_2}
-i{G_{00}+G_{33}\over\lambda_1}\bigg]\ \sigma_0
+\bigg[\langle F_3\rangle +{2iW_{03}^*\over\lambda_2}-{2iG_{03}\over\lambda_1}
\bigg]\ \sigma_3\cr
\equiv & -{2\pi\nu\over m_K}\left[\matrix{f_K & 0\cr
                                         0  & f_{\overline{K}}}\right]\ .
}
\eqno(3.5)
$$
The complex parameters $f_K$ and $f_{\overline{K}}$ now contain both the 
standard forward scattering amplitudes, described by $\langle F_0\rangle$ and
$\langle F_3\rangle$, and terms originating from the correlations (3.3).

The total effective hamiltonian $H$ can be expressed as in (2.7)
in terms of its eigenvalues $\lambda_S$, $\lambda_L$ and the quantities
$r_S$, $r_L$, $\theta$, $\sigma$ in matter, defined through its eigenvectors
as in (2.2), (2.8) and (2.9). The differences between these parameters and the corresponding
ones in vacuum can be expressed as a power series expansion in the combination:
$$
\eta={\pi\nu\over m_K}\bigg[{f_K-f_{\overline{K}}\over
\lambda_S^{(0)}-\lambda_L^{(0)} }\bigg]\ ;
\eqno(3.6)
$$
for many materials, one finds that $|\eta|$ is of order $10^{-2}\div10^{-3}$.[27, 28]
Although $\sigma=\sigma^{(0)}$ and therefore $T$ violation is unaffected by the medium,
to lowest order one obtains: $\theta=\theta^{(0)}-\eta$, so that,
not surprisingly, $CPT$ invariance is effectively broken by matter
effects even if it is preserved in vacuum; in addition,
one has: $r_S=r_S^{(0)}-\eta$, $r_L=r_L^{(0)}+\eta$.

\vskip .7cm

{\bf 3.2 Diagonal correlations}
\medskip

The noise hamiltonian $F(t)$ in (2.11) is no longer diagonal;
in this case the kaons undergo incoherent scatterings not exclusively
in the forward direction. However, we assume that the only non vanishing
correlation functions in the medium be diagonal, and precisely:
$$
\eqalign{
&\widehat{W}_{11}(t-s)=\widehat{W}_{00}(t-s)= W_{11}\ e^{-\lambda_1 |t-s|}\ ,\cr
&\widehat{W}_{22}(t-s)=\widehat{W}_{33}(t-s)= W_{22}\ e^{-\lambda_2 |t-s|}\ ,\cr
&\widehat{G}_{\mu\mu}(t-s)= G_{\mu\mu}\ e^{-\lambda_3 |t-s|}\ ,\quad \mu=0,1,2,3\ .
}
\eqno(3.7)
$$
The noise contribution to the effective hamiltonian takes now the form:
$$
H-H^{(0)}=\bigg[\langle F_0\rangle
+2i\bigg({W_{11}\over\lambda_1}+{W_{22}\over\lambda_2}\bigg)
-i\sum_{\mu=0}^3{G_{\mu\mu}\over\lambda_3}\bigg]\, \sigma_0
+\sum_{i=1}^3 \langle F_i\rangle\, \sigma_i\ ,
\eqno(3.8)
$$
while the matrix $\widehat{\cal C}_{ij}$ which characterizes the
dissipative term in (2.18) becomes diagonal:
$$
\widehat{\cal C}_{ij}=\left[\matrix{{2W_{11}\over\lambda_1} & &\cr
                             & {2W_{22}\over\lambda_2} & \cr
                             & & {2W_{22}\over\lambda_2}}\right]\ .
\eqno(3.9)
$$

\vskip .7cm

{\bf 3.3 General case}
\medskip

In general however, the fluctuations in the medium are
randomly directed so that the correlation functions in (2.12)
assume the generic form (3.3), with all components
of the constant matrices $G_{\mu\nu}$ and $W_{\mu\nu}$
non vanishing. In this case, keeping again only dominant contributions,
one finds:
$$
\widehat{\cal B}_{\mu\nu}={ G_{\mu\nu}\over\lambda_1}\ ,\qquad
\widehat{\cal C}_{\mu\nu}={2 W_{\mu\nu}\over\lambda_2}\ .
\eqno(3.10)
$$
As clear from the definitions $(2.12b)$ and (3.3), the matrix
$W$ turns out to be hermitian and also positive. The same 
properties clearly hold for the
\hbox{$3\times3$} coefficient submatrix $\widehat{\cal C}_{ij}$ 
characterizing the dissipative contribution in (2.18)
as given in (3.10), as well as for the previously
discussed cases (3.4) and (3.9). This makes the semigroup
${\mit\Gamma}_t:\ \hat\rho(0)\mapsto\hat\rho(t)={\mit\Gamma}_t[\hat\rho(0)]$
generated by the master equation (2.18) completely positive.
This property is crucial in assuring the consistency of
the generalized dynamics ${\mit\Gamma}_t$ 
in all possible situations;~\hbox{[1-3, 15, 29]}
as we shall see in discussing correlated neutral kaons, lacking of it
may lead to physically unacceptable results. It is therefore
reassuring that in our treatment the property of completely
positivity emerges naturally, without further assumptions.%
\footnote{$^{6)}$}{However, when the noise correlations assume
the more general exponential behaviour
$\widehat{W}_{\mu\nu}(t-s)=W_{\mu\nu}\ e^{-\lambda_{\mu\nu} |t-s|}$,
$\lambda_{\mu\nu}\geq0$, the corresponding matrix $\widehat{\cal C}_{ij}$ is no longer
automatically positive. In this case, complete positivity needs to be imposed
as an additional requirement.}

Although the matrix $\widehat{\cal C}_{ij}$ turns out to be in general 
complex, in the following we shall limit our considerations
only to situations for which $\widehat{\cal C}_{ij}$ results real.
These correspond to the physically most interesting cases;
in fact, the reality condition guarantees the increase of the von Neumann
entropy, $dS/dt\geq0$, $S=-{\rm Tr}(\hat\rho\ln\hat\rho)$,[29]
which is a physically desirable requirement for the dynamics of a small
system, the neutral kaon, in weak interaction with a large environment,
the medium.
For later convenience, we shall parametrize the entries of
$\widehat{\cal C}_{ij}$ using the real constants
$$
\eqalign{
&a=\widehat{\cal C}_{11}+\widehat{\cal C}_{22}\cr
&\alpha=\widehat{\cal C}_{11}+\widehat{\cal C}_{33}\cr
&\gamma=\widehat{\cal C}_{22}+\widehat{\cal C}_{33}
}
\hskip 2cm
\eqalign{
&b=\widehat{\cal C}_{23}\cr
&c=-\widehat{\cal C}_{13}\cr
&\beta=\widehat{\cal C}_{12}\ .
}
\eqno(3.11)
$$
The positivity condition on $\widehat{\cal C}_{ij}$ then requires:
$$
\eqalign{
&2\,R\equiv\alpha+\gamma-a\geq0\ ,\cr
&2\,S\equiv a+\gamma-\alpha\geq0\ ,\cr
&2\,T\equiv a+\alpha-\gamma\geq0\ ,\cr
&RST-2\, bc\beta-R\beta^2-S c^2-T b^2\geq 0\ .
}\hskip -1cm
\eqalign{
&RS-b^2\geq 0\ ,\cr
&RT-c^2\geq 0\ ,\cr
&ST-\beta^2\geq 0\ ,\cr
&\phantom{\beta^2}\cr
}\eqno(3.12)
$$

\vskip .7cm

{\bf 3.4 White noise}
\medskip

Finally, let us consider the case of a medium that behaves
as a white noise; it is described by stochastic variables
$F_\mu(t)$ that are $\delta$-correlated in time:
$$
\eqalignno{
&\widehat{G}_{\mu\nu}(t-s)=G_{\mu\nu}\ \delta(t-s)\ , &(3.13a)\cr
&\widehat{W}_{\mu\nu}(t-s)=W_{\mu\nu}\ \delta(t-s)\ , &(3.13b)
}
$$
with $G_{\mu\nu}$, $W_{\mu\nu}$ time-independent matrices,
such that $W=W^\dagger$. From the definition (2.20),
one immediately finds:
$$
\widehat{\cal B}_{\mu\nu}=G_{\mu\nu}\ ,\qquad
\widehat{\cal C}_{\mu\nu}=2W_{\mu\nu}\ ,
\eqno(3.14)
$$
without any approximation, and all considerations about the properties of
the generalized dynamics ${\mit\Gamma}_t$ presented in the
previous subsection can be repeated here. The present case is nevertheless
special: because of the correlations (3.13), all higher order terms in the expansion
(2.17) identically vanish, so that the evolution equation (2.18) is in
this case exact. 

\vskip 1cm

{\bf 4. OBSERVABLES}
\medskip

The evolution in time of the neutral kaons in the medium is fully described
by the reduced density matrix $\hat\rho(t)$, solution of the master equation
(2.18): any physical property of the system can be extracted from it 
by taking its trace with suitable hermitian operators.
Of particular interest are the observables $\widehat{\cal O}_f$ that are associated
with the decay of neutral kaons into a final state $f$, typically
$2\pi$, $3\pi$ and semileptonic $\pi\ell\nu$ states. In the fixed 
$\{|K^0\rangle,\ |\overline{K^0}\rangle\}$ basis, $\widehat{\cal O}_f$
can be represented by a $2\times2$ matrix, whose entries are expressed in
terms of the two independent decay amplitudes in vacuum
${\cal A}(K^0\rightarrow f)$, ${\cal A}(\overline{K^0}\rightarrow f)$:%
\footnote{$^{7)}$}{In the adopted treatment, kaon decays, being the result
of weak interaction processes, are unaffected by the presence of the media; 
all media influences are encoded in the dynamical equation (2.18).}
$$
\widehat{\cal O}_f=\left[\matrix{ |{\cal A}(K^0\rightarrow f)|^2 & 
\big[{\cal A}(K^0\rightarrow f)\big]^*\, {\cal A}(\overline{K^0}\rightarrow f)\cr
&\cr
{\cal A}(K^0\rightarrow f)\, \big[{\cal A}(\overline{K^0}\rightarrow f)\big]^* & 
|{\cal A}(\overline{K^0}\rightarrow f)|^2}\right]\ .
\eqno(4.1)
$$
The corresponding physical observable, directly accessible to the experiment,
is given by its mean value:
$$
\langle\widehat{\cal O}_f\rangle(t)=
{\rm Tr}\Big[\widehat{\cal O}_f\, \hat\rho(t)\Big]\ ,
\eqno(4.2)
$$
whose evolution in time is regulated by that of $\hat\rho(t)$.

Explicit solutions $\hat\rho(t)$ of the master equation (2.18) can be obtained
using perturbation theory. As discussed in the previous section, 
the entries of the coefficient
matrix $\widehat{\cal C}_{ij}$ characterizing the dissipative contribution
$\widehat{L}$ turn out to be of the form $W_{ij}/\lambda$, where $W_{ij}$
are the coefficients of the noise correlations, while $\lambda$ represents
their typical inverse decay time. From the analysis at the end of Section 2,
it follows that their magnitude can be evaluated to be
of order: $|\widehat{\cal C}_{ij}|\simeq 10^{-4}\ 
\Delta\Gamma^{(0)}(\Delta\Gamma^{(0)}/\lambda)$,
which is small, since by hypothesis $\lambda\gg\Delta\Gamma^{(0)}$.
These considerations allow treating the dissipative piece
$\widehat{L}$ in (2.18) as a perturbation to the contributions 
of the effective hamiltonian $H$,
which contains the standard Weisskopf-Wigner term $H^{(0)}$ of (2.7), 
besides the smaller,
higher order noise contributions $H^{(1)}$ and $H^{(2)}$.

It is convenient to work in a basis in which $H$ is diagonal:
all physical quantities (4.2), being the result of a trace operation, 
are basis independent. This can be obtained by a similarity transformation
analog to the one in (2.5):
$$
H=V\, H_D\, V^{-1}\ ,
\eqno(4.3)
$$
with
$$
V=\left[\matrix{ p_S & p_L \cr
                 q_S & -q_L}\right]\ ,\qquad
H_D=\left[\matrix{ \lambda_S & 0 \cr
                 0 & \lambda_L}\right]\ ,
\eqno(4.4)
$$
where, as before, $\lambda_S=m_S-i\gamma_S/2$, $\lambda_L=m_L-i\gamma_L/2$
are the two eigenvalues of $H$, while $p_S$, $p_L$, $q_S$, $q_L$
define the corresponding eigenvectors $|K_S\rangle$, $|K_L\rangle$
through definitions similar to those in (2.2).

This change of basis induces the transformation
$\hat\rho\rightarrow\rho=V^{-1} \hat\rho\, V^{-1\dagger}$
on the density matrix, while the master equation (2.18) becomes:
$$
{\partial \rho(t)\over\partial t}=-iH_D\, \rho(t) + i\rho(t)\, H_D^\dagger
+L[\rho(t)]\ ,
\eqno(4.5)
$$
where $L[\rho]$ is again of the form $(2.18c)$, but with a new coefficient
matrix ${\cal C}_{ij}$, whose entries are linear combinations of the
ones of the original matrix $\widehat{\cal C}_{ij}$ ({\it cf.} (3.11)).%
\footnote{$^{8)}$}{Explicitly, one has:
$L[\,\cdot\,]=V\ \widehat{L}\big[V\cdot V^\dagger\big]\,V^\dagger$;
however, since the coefficients $\widehat{\cal C}_{ij}$
in $\widehat{L}$ are small, in keeping with the approximation used before,
one can neglect $CP$ and $CPT$ violating effects in matter,
and set $p_S=p_L=q_S=q_L=1/\sqrt{2}$ in the above transformation law.}

Similarly, also the observables $\widehat{\cal O}_f$ get transformed,
$\widehat{\cal O}_f\rightarrow {\cal O}_f=V^\dagger\, \widehat{\cal O}_f\, V$, 
so that the corresponding mean value $\langle\widehat{\cal O}_f\rangle$
remains indeed unchanged:
$$
\langle\widehat{\cal O}_f\rangle\equiv\langle{\cal O}_f\rangle=
{\rm Tr}\Big[\rho(t)\, {\cal O}_f\Big]\ .
\eqno(4.6)
$$
The entries of the transformed matrix ${\cal O}_f$ can be conveniently
expressed in terms of the following two complex quantities:[17, 18]
$$
\lambda_S^f={q_S\over p_S}\, 
{{\cal A}(\overline{K^0}\rightarrow f)\over {\cal A}(K^0\rightarrow f)}\ ,\qquad
\lambda_L^f={q_L\over p_L}\, 
{{\cal A}(\overline{K^0}\rightarrow f)\over {\cal A}(K^0\rightarrow f)}\ ,
\eqno(4.7)
$$
as
$$
{\cal O}_f=\big|{\cal A}(K^0\rightarrow f)\big|^2\
\left[\matrix{|p_S|^2\,\big|1+\lambda_S^f\big|^2 & 
p_S^* p_L\,\big(1+\lambda_S^f\big)^*\, \big(1-\lambda_L^f\big)\cr
&\cr
              p_S p_L^*\, \big(1+\lambda_S^f\big)\, \big(1-\lambda_L^f\big)^*&
|p_L|^2\,\big|1-\lambda_L^f\big|^2}\right]\ ,
\eqno(4.8)
$$
or alternatively in terms of their reciprocal
$\mu_S^f=1/\lambda_S^f$, $\mu_L^f=1/\lambda_L^f$, so that
$$
{\cal O}_f=\big|{\cal A}(\overline{K^0}\rightarrow f)\big|^2\
\left[\matrix{|q_S|^2\,\big|1+\mu_S^f\big|^2 & 
-q_S^* q_L\,\big(1+\mu_S^f\big)^*\, \big(1-\mu_L^f\big)\cr
&\cr
            -q_S q_L^*\, \big(1+\mu_S^f\big)\, \big(1-\mu_L^f\big)^*&
|q_L|^2\,\big|1-\mu_L^f\big|^2}\right]\ .
\eqno(4.9)
$$

Using standard perturbation theory, solutions of (4.5) can then be easily 
obtained to any needed order in the small parameters ${\cal C}_{ij}$,
while keeping an exact dependence on the quantities parametrizing the
effective hamiltonian $H$, {\it i.e.} masses $m_S$, $m_L$ and widths
$\gamma_S$, $\gamma_L$ in matter, and the ratios
$r_S=p_S/q_S$, $r_L=p_L/q_L$. While these constants describe familiar
hamiltonian behaviour plus $CPT$ and $T$ violating effects in matter
[through the combinations $\theta$ and $\xi\equiv (|\sigma|-1)/(|\sigma|+1)$],
the additional piece $L$ in (4.5) is responsible for matter-induced non-standard
effects, leading to dissipation and quantum decoherence.
In the following we shall focus on the latter, analyzing in detail how 
the new effects modify the behaviour of selected neutral kaon observables.

\vskip 1cm

{\bf 5. SINGLE NEUTRAL KAONS}
\medskip

We shall start by discussing the time evolution of observables associated with
the propagation in random matter of single, uncorrelated neutral kaons;
typically,  these can be studied in fixed target experiments
({\it e.g.} see [30]).
As discussed in the previous section, relevant physical quantities
are the probability rates that an initial neutral kaon decays at a 
certain time $t$ into a given final state $f$:
$$
{\cal P}_f(K^0;t)={\rm Tr}\big[{\cal O}_f\, \rho_{K^0}(t)\big]\ ,\qquad
{\cal P}_f(\overline{K^0};t)={\rm Tr}\big[{\cal O}_f\, \rho_{\bar K^0}(t)\big]\ ,
\eqno(5.1)
$$
where $\rho_{K^0}(t)$, $\rho_{\bar K^0}(t)$ represent the evolution according
to (4.5) of  initially pure $K^0$,
$\overline{K^0}$ mesons states.

The case of semileptonic final states is of particular interest.
The amplitudes for the decay of a $K^0$ or
$\overline{K^0}$ state into $\pi^-\ell^+\nu$, $\pi^+\ell^-\bar\nu$
can be parametrized as follows:[31]
$$
\eqalignno{
&{\cal A}(K^0\rightarrow \pi^-\ell^+\nu)={\cal M} (1-y)\ , &(5.2a)\cr
&{\cal A}(\overline{K^0}\rightarrow \pi^+\ell^-\bar\nu)=
{\cal M}^* (1+y^*)\ , &(5.2b)\cr
&{\cal A}(K^0\rightarrow \pi^+\ell^-\bar\nu)= z\, 
{\cal A}(\overline{K^0}\rightarrow \pi^+\ell^-\bar\nu)\ , &(5.2c)\cr
&{\cal A}(\overline{K^0}\rightarrow \pi^-\ell^+\nu)=
x\, {\cal A}(K^0\rightarrow \pi^-\ell^+\nu)\ , &(5.2d)}
$$
where $\cal M$ is a common factor, while the parameters
$x$, $z$ measure violations of the $\Delta S=\Delta Q$ rule
and $y$ signals violations of the $CPT$-symmetry.
These quantities are expected to be very small, so that
one can consistently neglect terms containing $x$, $y$ or $z$
when multiplied by other small parameters, coming either
from the hamiltonian ({\it e.g.} $\theta$ and $\xi$)
or the dissipative part of the evolution equation (4.5).
In particular, this approximation implies that
for semileptonic decays: $\lambda^{\ell^+}_S=\lambda^{\ell^+}_L=x$,
$\mu^{\ell^-}_S=\mu^{\ell^-}_L=z$.

Then to first order in the dissipative matter effects, the probabilities (5.1)
explicitly read:%
\footnote{$^{9)}$}{Since $\Delta\Gamma\simeq 2\Delta m$ to a good approximation,
in writing (5.3) and (5.4) we have set the ratio $2\Delta m/\Delta\Gamma$
equal to one when multiplying small parameters.}
$$
\eqalignno{
{\cal P}_{\ell^+}(K^0;t)={|{\cal M}|^2\over2}\Bigg\{
&e^{-\Gamma t}\cos(\Delta m\, t)
\Bigg[{4\,{\cal R}e(r_S\, r_L^*)\over|r_S+r_L|^2}\,e^{-(A-D)\Delta\Gamma t}
-2\,{\cal R}e(y+2C)\Bigg]\cr
+&e^{-\Gamma t}\sin(\Delta m\, t)\Bigg[-{4\,{\cal I}m(r_S\, r_L^*)\over
|r_S+r_L|^2}
-2\,{\cal I}m(x)+{\cal R}e(B)\Bigg]\cr
+&e^{-\gamma_L t}\Bigg[{2|r_L|^2\over|r_S+r_L|^2}
-{\cal R}e(x+y -2C) +D 
\Bigg]\cr
+&e^{-\gamma_S t}\Bigg[{2|r_S|^2\over|r_S+r_L|^2}
+\,{\cal R}e(x-y+2C)-D\Bigg]\Bigg\}\ ,&(5.3a)\cr
&&\cr
&&\cr
{\cal P}_{\ell^-}(K^0;t)={|{\cal M}|^2\over2}\Bigg\{
&e^{-\Gamma t}\cos(\Delta m\, t)
\Bigg[-{4\, e^{-(A-D)\Delta\Gamma t}\over|r_S+r_L|^2}
-2\,{\cal R}e(y)\Bigg]\cr
+&e^{-\Gamma t}\sin(\Delta m\, t)\Bigg[2\,{\cal I}m(z+2C)-{\cal R}e(B)
\Bigg]\cr
+&e^{-\gamma_L t}\Bigg[{2\over|r_S+r_L|^2}
+{\cal R}e(y-z) -2\,{\cal I}m(C)+D\Bigg]\cr
+&e^{-\gamma_S t}\Bigg[{2\over|r_S+r_L|^2}+{\cal R}e(y+z)
+2\,{\cal I}m(C)-D\Bigg]\Bigg\}\ ,&(5.3b)\cr
}
$$
where $\Delta m$, $\Delta\Gamma$ and $\Gamma$ are defined as in (2.3)
in terms of masses and widths in medium. Matter induced dissipative
effects are controlled by the dimensionless parameters:
$$
A={\alpha+a\over\Delta\Gamma}\ ,\qquad
B={\alpha-a+2ib\over\Delta m}\ ,\qquad
C={c+i\beta\over\Delta\Gamma}\ ,\qquad
D={\gamma\over\Delta\Gamma}\ .
\eqno(5.4)
$$
The expressions for ${\cal P}_{\ell^+}(\overline{K^0};t)$ and 
${\cal P}_{\ell^-}(\overline{K^0};t)$ can be obtained from
$(5.3b)$ and $(5.3a)$ respectively, by changing the sign of $y$ and $C$,
and letting $r_S\rightarrow 1/r_S$, $r_L\rightarrow 1/r_L$,
$x\leftrightarrow z$.
For a non fluctuating medium, one has $A=B=C=D=0$ and the expressions in (5.3)
reduce to the standard ones, giving the probability of a semileptonic decay 
for a kaon that has travelled in a slab of material.

The probabilities ${\cal P}_\ell$ are directly accessible to the experiment.
Therefore with a suitable set-up, thanks to the different time dependence 
in the various pieces of (5.3), it is possible to extract information
on the parameters (5.4) and therefore on matter induced decoherence effects.
Clearly, this task becomes easier in the case of the generalized regeneration
discussed in Sect.3.1: in this case, the stochastic medium fluctuations are 
such that $A=D=B/2$, $C=0$.

Similar considerations apply to the study of the decay of a single neutral kaon
into $2\pi$ or $3\pi$ final states. For instance, in the just mentioned simplified case,
the $2\pi$-decay rate is sensible to the surviving dissipative constant $A$:
$$
\eqalign{
{\cal R}_{2\pi}(t)\equiv&{ {\rm Tr}\big[{\cal O}_{2\pi}\, \rho_{K^0}(t)\big]\over
{\rm Tr}\big[{\cal O}_{2\pi}\, \rho_{K^0}(0)\big] }
=e^{-\gamma_S t}\Bigg[ {|r_S|^2\, \big|1+\lambda_S^{2\pi}\big|^2\over
|r_S + r_L|^2 } -A\Bigg]\cr
&\hskip 1.2cm +e^{-\gamma_L t}\Bigg[ {|r_L|^2\, \big|1-\lambda_L^{2\pi}\big|^2\over
|r_S + r_L|^2 } +A\Bigg]
+2\, e^{-\Gamma t}\ |\eta_{2\pi}|\, \cos\big(\Delta m\, t-\phi_{2\pi}\big)\ ,}
\eqno(5.5)
$$
where
$$
\eta_{2\pi}\equiv |\eta_{2\pi}|\, e^{i\phi_{2\pi}}=
{r_S r_L^*\, \big(1+\lambda_S^{2\pi}\big)\big(1-\lambda_L^{2\pi}\big)^*
\over |r_S + r_L|^2 }\ ,
\eqno(5.6)
$$
while $\lambda_S^{2\pi}$, $\lambda_L^{2\pi}$ are the decay parameters defined
in (4.7). 

More in general, combining the semileptonic and pion decays probabilities
in suitable asymmetries one can obtain enough independent observables to treat
the case of more general media, for which the parameters in (5.4) are all
different from zero.

\vskip 1cm

{\bf 6. CORRELATED NEUTRAL KAONS}
\medskip

Matter induced dissipative effects can be further studied in experiments at
$\phi$-factories, using correlated neutral kaons.[19] Indeed, these set-ups behave like
quantum interferometers and therefore are particularly suitable for
analyzing phenomena leading to loss of quantum coherence.

In a $\phi$-factory, correlated kaons are produced from the decay of the
$\phi$ meson. Since this is a spin-1 particle, its decay into two spinless
bosons produces an antisymmetric spatial state. In the $\phi$ rest frame, the two
neutral kaons are then seen flying apart with opposite momenta,
and in the basis $|K^0\rangle$,
$|\overline{K^0}\rangle$, the resulting state can be described by:
$$
|\psi_A\rangle= {1\over\sqrt2}\Big(|K^0,-p\rangle \otimes  
|\overline{K^0},p\rangle -
|\overline{K^0},-p\rangle \otimes  |K^0,p\rangle\Big)\ .\eqno(6.1)
$$
The corresponding density operator $\hat\rho_A=|\psi_A\rangle\,\langle \psi_A|$ is
represented by a $4\times4$ matrix, since now it describes two kaons.
The time evolution of the correlated two kaon system can be expressed in terms
of the single meson dynamics ${\mit\Gamma}_t$ generated by the equation (2.18).
Indeed, once produced in a $\phi$-decay, the two kaons evolve independently,
so that the density matrix that describes the situation in which
the first kaon has evolved up to (proper) time $t_1$, while the second one
up to (proper) time $t_2$ is given by:
$$
\hat\rho_A(t_1,t_2)\equiv
\Big({\mit\Gamma}_{t_1}\otimes{\mit\Gamma}_{t_2}\Big)\big[\hat\rho_A\big]\ .
\eqno(6.2)
$$
Correspondingly, one can now study double decay observables,
{\it i.e.} the probability that a kaon decays into the final state
$f_1$ at time $t_1$, while the other kaon decays at time $t_2$
into the final state $f_2$:
$$
{\cal P}(f_1,t_1; f_2,t_2)=
\hbox{Tr}\Big[\Big(\widehat{\cal O}_{f_1}\otimes\widehat{\cal O}_{f_2}\Big) 
\ \hat\rho_A(t_1,t_2)\Big]\ ;
\eqno(6.3)
$$
here, $\widehat{\cal O}_{f_1}$, $\widehat{\cal O}_{f_2}$ 
are the $2\times2$ hermitian matrices
introduced in (4.1) that describe the decay of a single kaon
into the final states $f_1$, $f_2$, respectively.%
\footnote{$^{10)}$}{For the actual computation of the probabilities
${\cal P}(f_1,t_1; f_2,t_2)$, it is again convenient to work 
in the basis in which the effective hamiltonian
$H$ is diagonal, and therefore use
$\rho_A=\big[{V}^{-1}\otimes{ V}^{-1}\big]\ \hat\rho_A\
\big[{V}^{\dagger-1}\otimes{V}^{\dagger-1}\big]$,
and ${\cal O}_f$ in (4.8), (4.9).}

The probability rates in (6.3) are very sensitive to matter induced
decoherence effects. This is most strikingly shown by considering correlated decays
at equal time $t_1=t_2=t$ into the same final state $f_1=f_2=f$.
In absence of the dissipative term $\widehat{L}$ in the evolution equation (2.18),
the antisymmetry
properties of the initial state $\hat\rho_A$ would be preserved by the factorized
evolution ${\mit\Gamma}_{t}\otimes{\mit\Gamma}_{t}$, thus producing a vanishing
result for ${\cal P}(f,t; f,t)$. The equal time probabilities ${\cal P}(f,t; f,t)$
are therefore particularly suited to signal the presence of the dissipative
parameters in (5.4). For instance, for $f$ either a $2\pi$ or a $3\pi$ final state,
with associate intrinsic $CP$ parity $\zeta_f$, 
the probability ${\cal P}(f,t; f,t)$ is sensitive to the dissipative
parameter $D$; explicitly, one finds:
$$
{\cal P}(f,t; f,t)\propto 2D\Big( e^{-\gamma_L t} - e^{-\gamma_S t}\Big)\
\Big[(1+\zeta_f)e^{-\gamma_S t} + (1-\zeta_f)e^{-\gamma_L t}\Big]\ ,
\eqno(6.4)
$$
the proportionality constant being a decay amplitude normalization factor
dependenig on whether $f=2\pi$ or $3\pi$.
On the other hand, in the case of semileptonic final states, it is the
parameter $a$ that determines the slope at which the corresponding joint
probability approaches zero for small times; indeed, one has:
$$
{\cal P}(\ell^\pm,t; \ell^\pm,t)\sim a\, t\ .
\eqno(6.5)
$$

Although these probabilities together with the more general ones in (6.3)
can be measured at a $\phi$-factory,
much of the experimental analysis performed at these set-ups
is devoted to the study of integrated distributions
at fixed interval $t=t_1-t_2$:[32]
$$
{\mit\Gamma}(f_1,f_2;t)\equiv\int_0^\infty dt'\, {\cal P}(f_1,t'+t;f_2,t')\ ,\qquad
t\geq0\ .
\eqno(6.6)
$$
A particularly interesting observable that can be constructed with these
integrated probabilities involves $2\pi$ final states:
$$
{\cal A}_{\varepsilon^\prime}(t)={
{\mit\Gamma}(\pi^+\pi^-,2\pi^0;t) - 
{\mit\Gamma}(2\pi^0,\pi^+\pi^-;t)\over
{\mit\Gamma}(\pi^+\pi^-,2\pi^0;t) + 
{\mit\Gamma}(2\pi^0,\pi^+\pi^-;t) }\ ;
\eqno(6.7)
$$
it allows determining the ratio $\varepsilon^\prime/\varepsilon$,
where $\varepsilon$ and $\varepsilon^\prime$ are the familiar phenomenological
constants parametrizing the decay in two pions of the short and long-lived
kaons in vacuum.[17, 18] In a medium, the asymmetry ${\cal A}_{\varepsilon^\prime}(t)$
gets new contributions, both from the effective hamiltonian $H$
and the dissipative term of (2.18). To lowest order in all small parameters,
one explicitly finds:
$$
{\cal A}_{\varepsilon^\prime}(t)=3\, 
{\cal R}e\Big({\varepsilon^\prime\over\tilde\varepsilon}\Big)\ 
{N_1(t)\over N_3(t)}
-3\, {\cal I}m\Big({\varepsilon^\prime\over\tilde\varepsilon}\Big)\ 
{N_2(t)\over N_3(t)}\ ,\eqno(6.8)
$$
where
$$
\eqalign{
&N_1(t)=|\tilde\varepsilon|^2\Big( e^{-\gamma_L t} -  e^{-\gamma_S t}\Big)\ ,\cr
&N_2(t)=2\,|\tilde\varepsilon|^2\, e^{-\Gamma t}\sin(\Delta m\, t)\ ,\cr
&N_3(t)=e^{-\gamma_L t}\big(|\tilde\varepsilon|^2 + D\big)
+e^{-\gamma_S t}\bigg(|\tilde\varepsilon|^2-{\gamma_L\over\gamma_S} D\bigg)
-2\,|\tilde\varepsilon|^2\, e^{-\Gamma t}\cos(\Delta m\, t)\ ,
}
\eqno(6.9)
$$
and $\tilde\varepsilon=\varepsilon+\varepsilon_L-\varepsilon_L^{(0)}$,
with $\varepsilon_L=(r_L-1)/(r_L+1)$, 
$\varepsilon_L^{(0)}=(r_L^{(0)}-1)/(r_L^{(0)}+1)$.
A careful analysis of the time behaviour of the two contributions in
(6.8) would provide a way to estimate the dissipative parameter $D$,
together with real and imaginary part of 
$\varepsilon^\prime/\tilde\varepsilon$.
Further, note that in the long time limit the asymmetry (6.8) reduces to:
$$
{\cal A}_{\varepsilon^\prime}(\tau)\sim 3\, 
{\cal R}e\Big({\varepsilon^\prime\over\tilde\varepsilon}\Big)\
{ |\tilde\varepsilon|^2 \over |\tilde\varepsilon|^2 +D}\ ,
\eqno(6.10)
$$
and not simply to $3{\cal R}e\big(\varepsilon^\prime/\varepsilon\big)$,
as in vacuum. Therefore,
even assuming $\varepsilon_L\simeq\varepsilon_L^{(0)}$,
a measure of ${\cal A}_{\varepsilon^\prime}$ can no longer provide a
determination of ${\cal R}e(\varepsilon^\prime/\varepsilon)$
unless an estimate on the matter induced dissipative parameter $D$
is independently given.%
\footnote{$^{11)}$}{A non vanishing $D$ would decrease 
${\cal A}_{\varepsilon^\prime}$, making
${\cal R}e(\varepsilon^\prime/\varepsilon)$ bigger than measured.
In the case of string induced dissipative effects, this phenomenon have been
discussed in detail in [33]. Similar conclusions have also been mentioned in [25].}

\vskip 1cm

{\bf 7. DISCUSSION}
\medskip

Neutral kaons propagating in a stochastically fluctuating medium can be
treated as an open system, {\it i.e.} as a subsystem immersed in an
external environment. Starting from a microscopic hamiltonian
with a generic, linear kaon-matter interaction term, a generalized 
subdynamics for the kaon states has been explicitly derived
by averaging over the matter noise. It takes the form of a
completely positive quantum dynamical semigroup, where
the presence of the medium manifest itself through 
{\it i)} the generation
of hamiltonian corrections, that modify the familiar Weisskopf-Wigner 
description of the neutral kaon system, and {\it ii)}~the addition of extra 
pieces inducing dissipation and loss of quantum coherence.

Some of the hamiltonian contributions have been analyzed before in connection with
the so-called regeneration phenomena: they arise because of the coherent interaction
of the travelling kaons with the scattering centers in the medium.
On the contrary, the remaining hamiltonian pieces and 
the new dissipative contributions in the kaon evolution equation originate
from the stochastic correlations in the medium: as they move in the material,
the kaons encounter density fluctuations, whose correlations decay 
in time very rapidly with respect to the typical time scale
of the kaon system, thus inducing irreversibility and decoherence
in their dynamics.

Many physical phenomena can give rise to short time correlations
in ordinary materials: they have been studied by the so-called 
femtosecond chemistry.[24] By suitably inserting one of these materials
in any standard kaon-physics set-up, one can experimentally 
study the new, matter induced dissipative effects. Indeed,
as discussed in Sect. 5 and 6, one finds that both single and correlated kaon decay
observables get modified in a very specific way by the presence of these 
effects; as a result, they can be probed quite independently 
from other kaon physics phenomena.

Although here derived in a specific context,  
the generalized evolution equation (2.18) has wider validity: it has been shown to
generate the most general open system dynamics compatible
with a semigroup composition law and the requirement of complete positivity.[1-3]
As such, it has been recently applied to the description of dissipative effects
induced at low energies by the dynamics of fundamental objects (strings and branes)
at a very high scale, typically the Planck mass.[34] Although very small,
these string induced decoherence effects might be experimentally 
studied using interferometric devices,
like $\phi$-factories,[35, 36] and can in principle interfere with the
phenomena described here. Notice however that the two situations
correspond to totally different experimental conditions.
Matter generated phenomena are completely under the experimental control;
the effects they induce can easily be isolated from those that might be
generated at Planck scale by suitably varying the experimental conditions.

The possibility of choosing and tuning at will the experimental set-up 
further allows performing interesting tests on the physical consistency of the
dynamics generated by (2.18), and in particular on the property
of complete positivity.[29]
Consider the case in which only one of the two correlated kaons coming
from a $\phi$-meson decay actually propagates in a stochastic medium, while the
other evolve in vacuum. The density matrix that describes this situation at time $t$
is given by
$\tilde\rho_A(t,t)=\big({\mit\Gamma}_t\otimes{\mit\Gamma}_t^{(0)}\big)[\hat\rho_A]$,
where ${\mit\Gamma}_t$ is the map generated by the equation (2.18) evolving
the kaon in the medium, while ${\mit\Gamma}_t^{(0)}$, generated by the 
Weisskopf-Wigner hamiltonian $H^{(0)}$ of (2.7), describes the propagation
in vacuum of the second kaon.

As mentioned in Sect.2, any density matrix must be positive; this requirement
comes from the physical interpretation of its eigenvalues as probabilities,
that thus must be non-negative. In the case of a medium with diagonal
correlations as discussed in Sect.3.2, for which the parameters in (3.11)
are such that $a=\alpha$, $b=c=\beta=\,0$, the four eigenvalues
$\lambda_i(t)$, $i=1,2,3,4$, of the matrix
$\tilde\rho_A(t,t)$ above can be explicitly computed:
$$
\eqalign{
&\lambda_{1,2}(t)={\gamma\over\Delta\Gamma}\ e^{-i\gamma^{(0)}_{S,L} t}
\Big(e^{-i\gamma_L t} - e^{-i\gamma_S t}\Big)\ ,\cr
&\lambda_{3,4}(t)={1\over2}\Big\{\phi_1(t)+\phi_2(t)\pm
\Big[\big(\phi_1(t)-\phi_2(t)\big)^2 +4\,\psi_1(t)\psi_2(t)\Big]^{1/2}\Big\}\ ,
}
\eqno(7.1)
$$
where
$$
\eqalign{
&\phi_1(t)=e^{-\big[\gamma_S^{(0)}+\gamma_L\big]t}\cr
&\phi_2(t)=e^{-\big[\gamma_S+\gamma_L^{(0)}\big]t}
}
\qquad\quad
\eqalign{
&\psi_1(t)=e^{-\big[\Gamma_+ +\Gamma_-^{(0)} +2\alpha-\gamma\big]t}\cr
&\psi_2(t)=e^{-\big[\Gamma_+^{(0)} +\Gamma_- +2\alpha-\gamma\big]t}\ .
}
\eqno(7.2)
$$
Although $\lambda_1(t)$, $\lambda_2(t)$, $\lambda_3(t)$
are manifestly positive for any $t$, because of the minus sign in front 
of the square root, one can check that $\lambda_4(t)$ is non-negative
only for 
$$
2\alpha -\gamma\geq0\ .
\eqno(7.3)
$$ 
This is precisely the inequality that in this
case guarantees the condition of complete positivity of the dissipative evolution
${\mit\Gamma}_t$ (compare with (3.12)); lacking of it would have led
to physically inconsistent dynamics.

The situation just described can certainly be realized at
a $\phi$-factory, so that, at least in principle,
the time behaviour of the above eigenvalues can be experimentally studied,
and the inequality (7.3) probed. This would allow a direct test
of the condition of complete positivity, thus providing direct experimental
support for one of the crucial properties characterizing 
the quantum dynamics of open systems.

\vfill\eject

\centerline{\bf REFERENCES}
\vskip 1cm

\item{1.} R. Alicki and K. Lendi, {\it Quantum Dynamical Semigroups and 
Applications}, Lect. Notes Phys. {\bf 286}, (Springer-Verlag, Berlin, 1987)
\smallskip
\item{2.} V. Gorini, A. Frigerio, M. Verri, A. Kossakowski and
E.C.G. Surdarshan, Rep. Math. Phys. {\bf 13} (1978) 149 
\smallskip
\item{3.} H. Spohn, Rev. Mod. Phys. {\bf 52} (1980) 569
\smallskip
\item{4.} H.-P. Breuer and F. Petruccione, {\it The Theory of Open
Quantum Systems} (Oxford University Press, Oxford, 2002)
\smallskip
\item{5.} W.H. Louisell, {\it Quantum Statistical Properties of Radiation},
(Wiley, New York, 1973)
\smallskip
\item{6.} M.O. Scully and M.S. Zubairy, 
{\it Quantum Optics} (Cambridge University Press, Cambridge, 1997)
\smallskip
\item{7.} C.W. Gardiner and P. Zoller,
{\it Quantum Noise}, 2nd. ed. (Springer, Berlin, 2000)
\smallskip
\item{8.} R.R. Puri, {\it Mathematical Methods of Quantum Optics},
(Springer, Berlin, 2001)
\smallskip
\item{9.} {\it Atom Interferometry}, P.R. Berman, ed., (Academic Press,
New York, 1997)
\smallskip
\item{10.} P. Meystre, {\it Atom Optics}, (Springer-Verlag, New York, 
2001)
\smallskip
\item{11.} H. Rauch and S.A. Werner, {\it Neutron Interferometry}
(Oxford University Press, Oxford, 2000)
\smallskip
\item{12.} C.P. Slichter, {\it Principle of Magnetic Resonance}
(Springer-Verlag, Berlin, 1990)
\smallskip
\item{13.} C. Brosseau, {\it Fundamentals of Polarized Light},
(Wiley, New York, 1998)
\smallskip
\item{14.} F. Benatti and R. Floreanini, J. Optics B, {\bf 4} (2002) S238
\smallskip
\item{15.} F. Benatti, R. Floreanini and R. Romano, J. Phys. A
{\bf 35} (2002) 4955
\smallskip
\item{16.} F. Benatti and R. Floreanini, Phys. Rev. A {\bf 66} (2002)
043617
\smallskip
\item{17.} G.C. Branco, L. Lavoura and J.P. Silva, {\it CP Violation},
(Clarendon Press, Oxford, 1999)
\smallskip
\item{18.} I.I. Bigi and A.I. Silva, {\it CP Violation},
(Cambridge Univeristy Press, Cambridge, 2000)
\smallskip
\item{19.} {\it The Second Da$\,\mit\Phi$ne Physics Handbook}, 
L. Maiani, G. Pancheri and N. Paver, eds., (INFN, Frascati, 1995)
\smallskip
\item{20.} R.H. Good {\it et al.}, Phys. Rev. {\bf 124} (1961) 1223
\smallskip
\item{21.} P.K. Kabir, {\it The CP puzzle} (Academic Press, London, 1968)
\smallskip
\item{22.} P.H. Eberhard and F. Uchiyama, Nucl. Instr. Met. A {\bf 350} (1994) 144
\smallskip
\item{23.} R. Belusevic, {\it Neutral Kaons} (Springer-Verlag, Berlin, 1999)
\smallskip
\item{24.} A.H. Zewail, J. Phys. Chem. A {\bf 104} (2000) 5660
\smallskip
\item{25.} A.A. Andrianov, J. Taron and R. Tarrach, 
Phys. Lett. {\bf B507} (2001) 200
\smallskip
\item{26.} J. Budimir and J.L. Skinner, J. Stat. Phys. {\bf 49} (1987) 1029
\smallskip
\item{27.} A. Di Domenico, Nucl. Phys. {\bf B450} (1995) 293
\smallskip
\item{28.} Particle Data Group, Phys. Rev. D {\bf 66} (2002) 010001
\smallskip
\item{29.} F. Benatti and R. Floreanini,
Mod. Phys. Lett. {\bf A12} (1997) 1465;
Banach Center Publications, {\bf 43} (1998) 71;
Phys. Lett. {\bf B468} (1999) 287; Chaos, Solitons and Fractals
{\bf 12} (2001) 2631
\smallskip
\item{30.} CPLEAR Collaboration, Phys. Rep. {\bf 374} (2003) 165
\smallskip
\item{31.} N.W. Tanner and R.H. Dalitz, Ann. of Phys. {\bf 171} (1986) 463
\smallskip
\item{32.} G. D'Ambrosio, G. Isidori and A. Pugliese, $CP$ and $CPT$
measurements at Da$\Phi$ne, in Ref.[19]
\smallskip
\item{33.} F. Benatti and R. Floreanini, Mod. Phys. Lett. {\bf A14}
(1999) 1519
\smallskip
\item{34.} F. Benatti and R. Floreanini, Ann. of Phys. {\bf 273} (1999) 58
\smallskip
\item{35.} F. Benatti and R. Floreanini, Nucl. Phys. {\bf B511} (1998) 550
\smallskip
\item{36.} F. Benatti, R. Floreanini and R. Romano, Nucl. Phys. {\bf B602}
(2001) 541

\bye